\documentclass{elsart}

\usepackage{natbib}

\usepackage{epsfig}

\usepackage{amssymb}

\begin{document}

\begin{frontmatter}


 \title{A cross-correlation of WMAP and ROSAT}
 \author{J.M. Diego$^1$, J. Silk$^1$, W. Sliwa$^2$}
\address{$^1$University of Oxford, Denys Wilkinson Building, 1 Keble Road, 
            OX1 3RH Oxford, United Kingdom}
\address{$^2$Nicolas Copernicus Astronomical Centre, Bartycka 18, 
             00-716 Warsaw, Poland}



\author{}

\address{}

\begin{abstract}
We cross-correlate the recent CMB WMAP 1 year data with the diffuse 
soft X-ray background map of ROSAT. We look for common signatures due to galaxy 
clusters (SZ effect in CMB, bremsstrahlung in X-rays) by cross-correlating the two 
maps in real and in Fourier space. We do not find any significant correlation 
and we explore the different reasons for this lack of correlation. The most 
likely candidates are the possibility that we live in a low $\sigma _8$ 
universe ($\sigma _8 < 0.9$) and/or systematic effects in the data especially 
in the diffuse X-ray maps which may suffer from significant cluster signal 
subtraction during the point source removal process. 
\end{abstract}


\end{frontmatter}

\section{Introduction}
\label{sect_intro}
The recent release of the WMAP data (Bennett et al. 2003) has opened a new window for 
studies of large-scale structure based on the well known Sunyaev-Zel'dovich 
effect (SZ effect) (Sunyaev \& Zel'dovich, 1972).
The SZ effect shifts the spectrum of the CMB photons to higher frequencies. 
This shift is redshift-independent and proportional to the product
of the electron column density with the {\it average} temperature along the line of 
sight. The electron temperature and optical depth due to Thomson scattering are 
particularly high inside galaxy clusters. 
Thus, the SZ effect is a good tracer of clusters, even for those at 
high redshift. Around galaxy clusters,  a diffuse, possibly filamentary,  
distribution of hot gas is believed to be present. These filaments have not been 
definitively detected due to their low contrast compared with the background 
(either CMB or X-ray backgrounds). 
The same electrons which cause the SZ effect will also
emit X-rays by bremsstrahlung emission. Therefore, one expects the SZ effect and 
the X-ray emission of galaxy clusters and filaments to be spatially correlated. 
Since the X-ray background and the CMB are not correlated (except at very 
large scales where there could be a correlation due to the integrated Sachs-Wolfe 
effect (Boughn et al. 1998)), the  cross-correlation of an X-ray map with the CMB should 
enhance the signal of clusters and filaments with respect to the
background. This fact motivates the present study.\\

We will be interested in studying the cross-correlation 
$SZ\otimes XR$ (where $\otimes$ stands for cross-correlation). 
We need to define a {\it statistical object} to quantify this 
correlation. We will use the power spectrum of the $SZ\otimes XR$ map 
as such an object. We will also use the so-called 
cross-power spectrum (cross-correlation of the Fourier modes). 
There are several advantages to using the power 
spectrum and cross-power spectrum over other statistical objects. 
First, they contain useful information at different scales. For instance the 
0 mode accounts for the correlation coefficient of the two maps. 
Higher modes will contain information about the fluctuations at 
smaller scales. The modelling of the power spectrum is 
also easier and it can easily account for the uncertainties in the assumptions 
made in the model. 
The power spectrum will also tell us something about the contribution of 
clusters and filaments to the CMB power spectrum. Previous papers have claimed 
an excess in the CMB power spectrum (Mason et al. 2003; Bond et al. 2002). 
It is not yet clear whether  this excess could be caused by the SZ effect signal or just
be inadequately subtracted residuals (compact sources or residual noise). 
An independent estimation of the SZ effect power spectrum would help to clarify 
this point. 
The reader can find all the details of this work on Diego et al. (2003)

\section{The cross-correlation WMAP$\otimes$ROSAT}\label{section_cross}
WMAP data consists of 5 all-sky maps at five different frequencies 
(23 Ghz $<$ $\nu$ $<$ 94 Ghz). At low frequencies, these maps show strong 
galactic emission (synchrotron and free-free). The highest frequency maps 
(41-94 Ghz) are the cleanest in terms of galactic contaminants and will be 
the most interesting for our purpose.
We will focus on one basic linear combination of the WMAP data, the differenced 
$Q-W$ bands of the $1^{\circ}$ smoothed version of the original data. 
This differencing completely removes the main contaminant in this work, the CMB,  
leaving a residual dominated by galactic and extragalactic foregrounds as well  
as filtered instrumental noise.\\
On the other hand, the ROSAT All-Sky Survey data (RASS, see Snowden et al. 1997) 
is presented in a set of bands ($\approx 0.1-2$ keV). Low energy bands are 
highly contaminated by local emission (local bubble and Milky Way galaxy) while high 
energy bands show an important contribution from extragalactic AGN's. 
The optimal band for our purposes will be the band R6 ($\approx$ 0.9-1.3 keV). 
This band is the best in terms of instrumental response, background contamination 
and cluster vs AGN emission. 
The ROSAT maps have been {\it cleaned} from the most prominent 
point sources (sources above the threshold 0.02 cts/s in the R5+R6 band). 
However, we should note that for the above threshold, the survey source catalogue was 
incomplete and there are still some (very hard) point sources present 
in the maps of the diffuse X-ray background. We also have to note that during 
the point source subtraction process several {\it compact} clusters were removed 
from the data. This fact may introduce a systematic error in our conclusions which 
will be discussed later.\\

In order to maximise the extragalactic signal, 
we  restrict our analysis to regions outside the galactic plane. 
In particular, we will consider only a {\it clean} portion of the sky 
above $b=40^{\circ}$ and $70^{\circ} < \ell < 250^{\circ}$ which will
also exclude  the contribution from the north-galactic spur. 
This {\it optimal} area of the sky covers $\approx 9 \%$ of the sky. \\
We also have to remove a few bright point sources in the ROSAT maps which were 
not originally removed.
Some of these sources will produce a correlation signal if we do not remove 
them. One of these bright sources (MRK 0421) was already identified in a previous 
work as a source of correlation between ROSAT and COBE maps (Kneissl et al. 1997).

\begin{figure}
   \begin{center}
   \epsfysize=6.cm 
   \begin{minipage}{\epsfysize}\epsffile{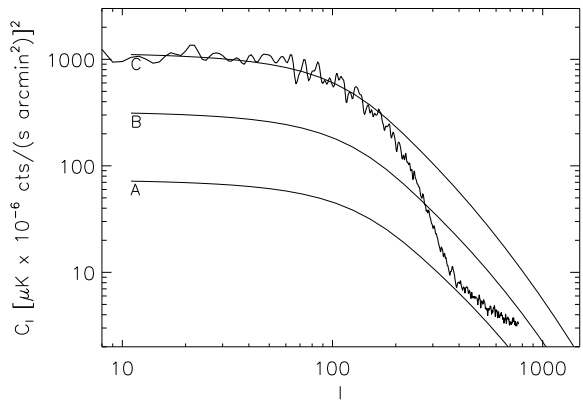}
   \end{minipage}
   \caption{Power spectrum of (Q-W)WMAP times ROSAT-R6. 
            The thin solid lines are the predicted signals for models with  
            $\sigma _8 = 0.8$ (A), $\sigma _8 = 0.9$ (B) and 
            $\sigma _8 = 1.0$ (C). 
           }
   \label{fig_power_3models}
   \end{center}
\end{figure}

The power spectrum of $WMAP\otimes ROSAT$ (power spectrum of the product of 
the maps in real space) is shown in figure \ref{fig_power_3models}. 
At large scales (small multipoles) the product maps contains more power that at 
small scales (large $l$'s). This is mostly due to the smoothing process of WMAP maps 
($1^{\circ}$) which suppresses power at small scales. We should also mention that,  
contrary to what happens in the cross-power spectrum, the power spectrum of the 
product maps does not have to be 0 if there is no correlation between the maps. 
This fact makes more difficult the identification of a correlation between two maps. 
In order to identify possible correlations we have to rotate one of the maps. Then, 
if there is a spatial correlation between the maps, the power spectrum will be smaller 
after rotating one of the maps. When we do that for several rotation angles we observe 
that the power spectrum does not change. This means that there is not a significant 
correlation between the maps. 
 
When we look at the cross-power spectrum, we do not observe any significant correlation 
either. The cross-power spectrum oscillates around 0 as expected for two maps which are 
not correlated. The cross-power spectrum is the $k$-ring average of the product of the 
Fourier modes of the two maps. If the maps are not correlated this average must be 0. \\

\begin{figure}
   \begin{center}
   \epsfysize=6.cm 
   \begin{minipage}{\epsfysize}\epsffile{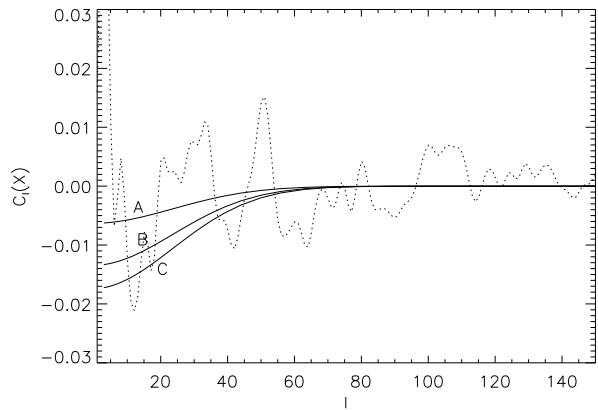}
   \end{minipage}
   \caption{
            Cross-power spectrum (dotted line) compared with the models 
            $\sigma _8 = 0.8$ (A), $\sigma _8 = 0.9$ (B) and 
            $\sigma _8 = 1.0$ (C). 
           }
   \label{fig_crosspower2}
   \end{center}
\end{figure}

Although we do not detect any signal neither with the power spectrum of the product maps 
nor the cross-power spectrum, we can still use this fact to set some constraints on the model.
In figures \ref{fig_power_3models} and \ref{fig_crosspower2} we compare the measured 
power and cross-power with three different models where we change the parameter $\sigma _8$ 
($\Omega _m$ is fixed to 0.3). This simple comparison tell us that 
models with low $\sigma _8$ are favoured by the observed lack of correlation between the 
two data sets. 
High values of $\sigma _8$ can be accommodated if the SZ effect is significantly 
contaminated by point sources (embedded in the clusters) so the net distortion in the 
CMB is smaller than if the cluster signal is just pure SZ effect 
(point sources contribute as positive signals while the SZ effect is negative at the 
frequencies considered in this work). 
High values of $\sigma _8$ can also be compatible with the data if the fraction of clusters 
removed in the point source subtraction process in ROSAT data contribute significantly to 
the power spectrum and cross-power spectrum. We have estimated this possible source in 
systematic error in the {\it worst-scenario} case in which all the clusters above the 
0.02 cts/s point source removal threshold have been subtracted. In this case the 
theoretical power spectrum shown in figures \ref{fig_power_3models} and \ref{fig_crosspower2} 
should be smaller by a factor $\approx 3$. Taking this reduction in power into account, models 
with $\sigma _8 = 0.9$ could be marginally consistent with the lack of correlation but 
models with $\sigma _8 = 1.0$ still seem difficult to reconcile.

\section{conclusions}
We do not find any significant correlation between WMAP and ROSAT. This lack 
of correlation can be due to contamination of the SZ effect by point sources, 
a significant reduction in the cluster signal due to erroneously subtracted clusters 
during the point source subtraction process in ROSAT data, the possibility that we live 
in a universe with a low normalisation of the power spectrum $\sigma _8$ or a combination of 
the previous factors.\\
We found that different assumptions about the model lead to different 
fits to the data. In particular, high values of $\sigma _8$ seem to be difficult 
to reconcile with the absence of significant correlation. \\
\begin{figure}
   \begin{center}
   \epsfysize=6.cm 
   \begin{minipage}{\epsfysize}\epsffile{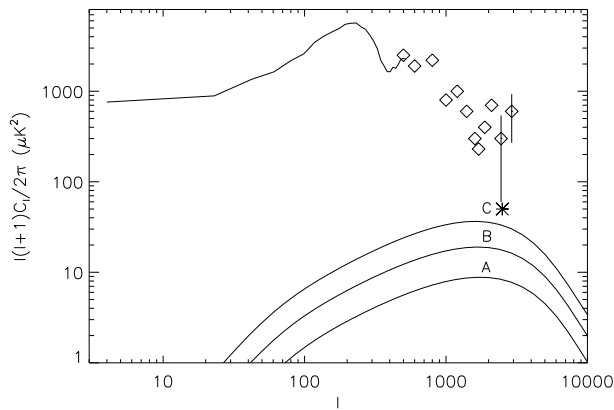}
   \end{minipage}
   \caption{Current estimates of the CMB power spectrum compared with 
            predicted SZ effect power spectrum (R-J) for the previous 
            models A,B, and C
            The top solid line is a rebinning (10 bins) of the 
            original WMAP CMB power spectrum. The symbols are current estimates by CBI and 
            ACBAR (error bars have been omitted except in the last two points). 
            The last three symbols at $\ell \approx 3000$ 
            are the estimated power spectrum at high $\ell$ by CBI (top), ACBAR (middle) 
            and the expected CMB power spectrum for a standard model (bottom star).
            }
   \label{fig_Cl_WMAP_vs_SZ}
   \end{center}
\end{figure}
This absence of significant correlation could also be used to rule out the possibility 
that the excess in ACBAR and CBI is due to SZ effect. We illustrate this point 
in figure \ref{fig_Cl_WMAP_vs_SZ} where we compare the power spectrum of the SZ effect for 
the models A,B, and C with the recent estimate of the CMB power spectrum by 
WMAP (solid line) and with estimates from ACBAR (Kuo et al. 2003) 
and CBI (Mason et al. 2003). \\
From this plot we could conclude that the fact that we do not observe a correlation between 
WMAP and ROSAT implies that the SZ effect power spectrum should not contribute significantly 
to any of these experiments. Unfortunately, the quality of the data does not allow us to make 
such a severe affirmation. An updated version of this work made with the four year CMB data 
from WMAP and a new version of the soft X-ray diffuse background maps (with no cluster 
subtraction) could certainly help to answer this question.

\section{Acknowledgements}
This research has been supported by a Marie Curie Fellowship 
of the European Community programme {\it Improving the Human Research 
Potential and Socio-Economic knowledge} under 
contract number HPMF-CT-2000-00967. 







\begin{thebibliography}{}


\bibitem{Bennett2003} Bennet et al. 2003, ApJ accepted, astro-ph/0302207

\bibitem{Bond2002} Bond J.R., et al. 2002, ApJ submitted.
                   astro-ph/0205386.
\bibitem{Boughn98} Boughn et al. 1998, Boughn S.P., Crittenden R.G., Turok N., 
                   1998, New Astronomy, vol 3, no 5, p. 275.

\bibitem{Diego2003} Diego J.M., Silk J.  W. Sliwa, 2003, MNRAS accepted. astro-ph/0302268

\bibitem{Kneissl1997} Kneissl R., Egger R., Hasinger G., Soltan A.M., Tr\"{u}mper J., 
                      1997, A\&A, 320, 685.

\bibitem{Kuo2003} Kuo et al. 2003, ApJ submitted. astro-ph/0212289.

\bibitem{Mason2003} Mason et al. 2003, ApJ, 591, 540.

\bibitem{Snowden1997} Snowden S.L., Egger R., Freyberg M.J., McCammon D., Plucinsky P.P.,
                      Sanders W.T., Schmitt,J. H.M.M., Truemper J., Voges W.
                      1997, ApJ, 485, 125.

\bibitem{Sunyaev72} Sunyaev R.A., Zel'dovich Ya, B., 1972, A\&A, 20, 189.

\end{thebibliography}
\end{document}